\documentclass[twocolumn,preprintnumbers,amsmath,amssymb]{revtex4}

\usepackage{mathrsfs}%====================================
\usepackage{amssymb}
\usepackage{graphicx}% Include figure files
\usepackage{dcolumn}% Align table columns on decimal point
\usepackage{bm}% bold math
\usepackage{textcomp}
\usepackage{amsmath}
 \usepackage{color}
\usepackage[titletoc]{appendix}
\usepackage{multirow}
\usepackage{diagbox}

\newcommand\ket[1]{\ensuremath{|#1\rangle}}

\newcommand\oprod[2]{\ensuremath{|#1\rangle\langle#2|}}
\newcommand\mean[1]{\ensuremath{\langle #1\rangle}}

\newcounter{RomanNumber}

\begin{document}
\title{Higher key rate of measurement-device-independent quantum key distribution through joint data processing}
\author{Cong Jiang$ ^{1}$, Zong-Wen Yu$ ^{2,4}$, Xiao-Long Hu$ ^{2}$, 
and Xiang-Bin Wang$ ^{1,2,3,5\footnote{Email Address: xbwang@mail.tsinghua.edu.cn}\footnote{Also at Center for Atomic and Molecular Nanosciences, Tsinghua University, Beijing 100084, China}}$}

\affiliation{ 
\centerline{$^{1}$ Jinan Institute of Quantum technology, SAICT, Jinan 250101, China}
\centerline{$^{2}$State Key Laboratory of Low
Dimensional Quantum Physics, Department of Physics,} \centerline{Tsinghua University, Beijing 100084, China}
\centerline{$^{3}$ Synergetic Innovation Center of Quantum Information and Quantum Physics,}\centerline{University of Science and Technology of China, Hefei, Anhui 230026, China}
\centerline{$^{4}$Data Communication Science and Technology Research Institute, Beijing 100191, China}
\centerline{$^{5}$ Shenzhen Institute for Quantum Science and Engineering, and Physics Department,}
\centerline{Southern University of Science and Technology, Shenzhen 518055, China}}
%%%%%%%%%%%%%%%%%%%%%%%%%%%%%%%%%%%%%%%%%%%%%%%%%%%%%%%%%%%%%%%%%%%
%%%%%%%%%%%%%%%%%%%%%%%%%%%%%%%%%%%%%%%%%%%%%%%%%%%%%%%%%%%%%%%%%%%
%%%%%%%%%%%%%%%%%%%%%%%%% Abstract %%%%%%%%%%%%%%%%%%%%%%%%%%%%%%%%
\begin{abstract}
We propose a method named as double-scanning method, to improve the key rate of measurement-device-independent quantum key distribution (MDI-QKD) drastically. In the method, two parameters are scanned simultaneously to tightly estimate the counts of single-photon pairs and the phase-flip error rate jointly. Numerical results show that the method in this work can improve the key rate by $35\%-280\%$ in a typical experimental set-up. Besides, we study the optimization of MDI-QKD protocol with all parameters including the source parameters and failure probability parameters, over symmetric channel or asymmetric channel. Compared with the optimized results with only the source parameters, the all-parameter-optimization method could improve the key rate by about $10\%$.
\end{abstract}

%%%%%%%%%%%%%%%%%%%%%%%%%%%%%%%%%%%%%%%%%%%%%%%%%%%%%%%%%%%%%%%%%%%
%%%%%%%%%%%%%%%%%%%%%%%%%%%%%%%%%%%%%%%%%%%%%%%%%%%%%%%%%%%%%%%%%%%
%%%%%%%%%%%%%%%%%%%%%%%%%%%%%%%%%%%%%%%%%%%%%%%%%%%%%%%%%%%%%%%%%%%

\maketitle
\section{Introduction}
The first quantum key distribution (QKD) protocol, BB84 protocol~\cite{bennett1984quantum} is proposed by Bennett and Brassard in 1984. Based on the quantum laws, QKD could provide unconditionally secure private communication between two parties, Alice and Bob~\cite{gisin2002quantum,gisin2007quantum,xu2020secure,pirandola2019advances,
scarani2009security,shor2000simple}. But the security of the original BB84 protocol is under the assumption of perfect single photon sources, or else its security would be destroyed by photon number splitting (PNS) attack~\cite{huttner1995quantum,brassard2000limitations}. The decoy-state method~\cite{hwang2003quantum,wang2005beating,lo2005decoy} is proposed to assure the security of BB84 protocol with imperfect single photon sources such as weak coherent state (WCS) sources. The decoy-state BB84 protocol greatly improves the key rate and secure QKD distance in practice and has been widely studied in theory~\cite{wang2007quantum,ad2007simple,wang2007simple,wang2008general,scarani2008quantum,wang2009decoy,
hayashi2012concise,tomamichel2012tight,lim2014concise,
tamaki2014loss,yu2016reexamination,chau2018decoy}. Many experiments of decoy-state BB84 protocol have been reported~\cite{rosenberg2007long,schmitt2007experimental,peng2007experimental,boaron2018secure,wang2008experimental}. And the farthest secure QKD distance of BB84 protocol in fiber reaches up to 421 km~\cite{boaron2018secure}. The decoy-state BB84 protocol is also applied to QKD between ground and satellite~\cite{liao2017satellite} and QKD networks~\cite{peev2009secoqc,chen2010metropolitan,
sasaki2011field}. Besides decoy-state mehtod, the round-robin differential-phase-shift protocol can also effectively defence the PNS attack~\cite{sasaki2014practical,takesue2015experimental}.

Besides the imperfect single photon sources, the imperfect detectors in Bob's laboratory can also be attacked by Eve~\cite{lydersen2010hacking,gerhardt2011full}. Measurement-Device-Independent (MDI)-QKD~\cite{braunstein2012side,lo2012measurement} protocol was proposed to solve all possible detection loopholes. The security of decoy-state MDI-QKD protocol with imperfect sources and detectors has been proved in both infinite key size~\cite{lo2012measurement} and finite key size~\cite{curty2014finite}. Many improved schemes of decoy-state MDI-QKD protocol have been proposed to improve the key rate~\cite{tamaki2012phase,wang2013three,xu2013practical,xu2014protocol,yu2015statistical,zhou2016making,hu2017practical} and assure its security in practice~\cite{jiang2016measurement,jiang2017measurement}. The theories of decoy-state MDI-QKD protocol have been widely demonstrated in experiments~\cite{rubenok2013real,liu2013experimental,tang2014experimental,wang2015phase,comandar2016quantum,
yin2016measurement,wang2017measurement,pirandola2015high,semenenko2020chip,cao2020chip}. Among all those theories and experiments, the 4-intensity MDI-QKD protocol~\cite{zhou2016making} performs the best and has been the mainstream protocol of MDI-QKD. Our 4-intensity MDI-QKD protocol has been applied successfully in a number of important experiments: the long distance MDI-QKD over 404 km~\cite{yin2016measurement}, the high rate MDI-QKD experiment~\cite{pirandola2015high}, the fault-tolerant MDI-QKD experiment~\cite{wang2017measurement}, the on-chip MDI-QKD system ~\cite{semenenko2020chip,cao2020all}, and very recently, the free-space MDI-QKD~\cite{cao2020long}.  In theoretical studies, the 4-intensity has been further studied for the asymetric channel~\cite{wang2019asymmetric,hu2018measurement} which is useful for a network QKD~\cite{wang2019asymmetric} and the unstable channel~\cite{hu2018measurement} which is useful for the free-space QKD. Very recently, it is studied with new statistical inequalities \cite{chau2020security} to improve the performance.

In the original 4-intensity MDI-QKD protocol~\cite{zhou2016making}, an important idea is to consider the constraints jointly. Here, we add new joint constraints with a double-parameter scan: we simultaneously scan the error counts and the vacuum related counts and get the worst-case jointly for the counting rate of single-photon pulses and phase-flip error rate. Since new constraints are added, the key rate is improved drastically. 

{In our prior art work~\cite{zhou2016making}, we take global optimization with source parameters including the intensities of light sources and their corresponding sending probabilities.} In this paper, we propose a double-scanning method of the 4-intensity MDI-QKD, and study the global optimization of the 4-intensity MDI-QKD protocol with finite-size effect. The optimized parameters include not only the intensities of light sources and their corresponding sending probabilities, but also the tens of failure probabilities in the finite-size effect analysis. Here we propose a new optimization method to complete the task with so many parameters. We use Chernoff bound~\cite{chernoff1952measure} for our calculation. Based on the method proposed here, we simulate the key rate of the 4-intensity MDI-QKD protocol with symmetric and asymmetric channels.

\section{Set-up of the 4-intensity MDI-QKD protocol}
\begin{figure}
\centering
\includegraphics[width=8cm]{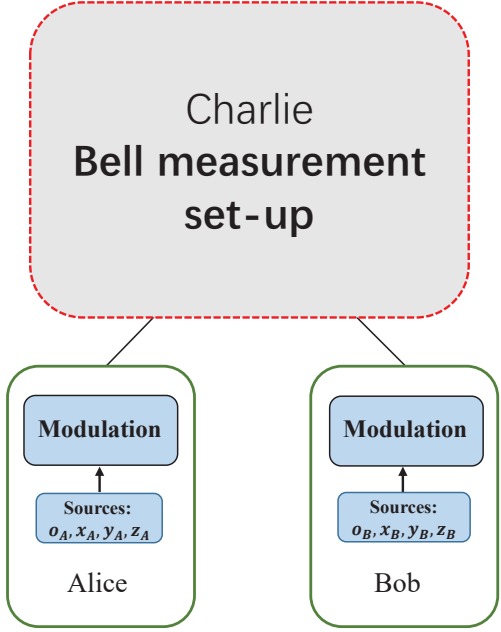}
\caption{{The schematic set-up of 4-intensity MDI-QKD protocol~\cite{zhou2016making}. Alice (Bob) takes encoding in $X$ basis for sources $o_A,x_A,y_A$ ($o_B, x_B, y_B$) and encoding in $Z$ basis for source $z_A$ ($z_B$). In practice, there is no complete Bell measurement device for Charlie. Even in such a case, Alice and Bob can still use those effective events for the decoy-state analysis and final key distillation~\cite{lo2012measurement,zhou2016making}.}}\label{model}
\end{figure}

In the 4-intensity MDI-QKD protocol~\cite{zhou2016making}, there are four different intensities of sources at Alice's and Bob's sides respectively. {In the $X$ basis, Alice (Bob) uses source $o_A, x_A$, or $y_A$ ($o_B, x_B$, or $y_B$) and takes bit value encoding in the $X$ basis. In the $Z$ basis, Alice (Bob) uses source $z_A$ ($z_B$) and takes bit value encoding in the $Z$ basis. Each side takes BB84 encoding in $Z$ basis $\{\ket{0}, \ket{1}\}$ and $X$ basis $\{\ket{\pm}=\frac{1}{\sqrt{2}}(\ket{0}\pm\ket{1})\}$. The theory of the 4-intensity protocol applies to any specific physical realization of state encoding, e.g., the polarization encoding, the phase encoding, the time-bin encoding, and so on.} The intensities of Alice need not to be the same with those of Bob, e.g., in the situation of asymmetric channel shown in Ref.~\cite{hu2018measurement}. In the whole protocol, Alice and Bob send $N$ pulse pairs to Charlie. 

{In the photon-number space, the density matrices of the sources of Alice and Bob are
\begin{equation*}
\rho_{l_A}=\sum_{j}a_j^l\oprod{j}{j},\rho_{r_B}=\sum_{k}b_k^r\oprod{k}{k},l,r=o,x,y,z.
\end{equation*} 
And we assume 
\begin{equation}\label{const}
\frac{a_j^y}{a_j^x}\ge \frac{a_2^y}{a_2^x}\ge \frac{a_1^y}{a_1^x}, \frac{b_k^y}{b_k^x}\ge \frac{b_2^y}{b_2^x}\ge \frac{b_1^y}{b_1^x},
\end{equation}
hold for any $j>2$ and $k>2$. In this paper, we take the phase-randomized WCS sources as an example to show our calculation method and numerical results, but both our double-scanning method proposed here and the single-scanning method proposed in Ref.~\cite{zhou2016making} can be applied to all sources that satisfy Eq.~\eqref{const}, such as the heralded single-photon sources.}

In the $i$th time window, as shown in Figure~\ref{model}, Alice (Bob) prepares a phase-randomized WCS pulse whose intensity is randomly chosen from $\mu_{o_A}=0$, $\mu_{x_A}$, $\mu_{y_A}$, or $\mu_{z_A}$ for sources $o_A, x_A,y_A$ or $z_A$ ($\mu_{o_B}=0$, $\mu_{x_B}$, $\mu_{y_B}$, or $\mu_{z_B}$ for sources $o_B, x_B,y_B$ or $z_B$) with probability $p_{o_A}=1-p_{x_A}-p_{y_A}-p_{z_A}$, $p_{x_A}$, $p_{y_A}$, and $p_{z_A}$ ($p_{o_B}=1-p_{x_B}-p_{y_B}-p_{z_B}$, $p_{x_B}$, $p_{y_B}$, and $p_{z_B}$) respectively. Here, the constraints Eq.~\eqref{const} are equivalent to
\begin{equation*}
\mu_{x_A}\le\mu_{y_A},\quad \mu_{x_B}\le\mu_{y_B}.
\end{equation*} 
As the phases of Alice's and Bob's phase-randomized WCS pulse are never announced, those pulses are actually the classical mixture of different photon numbers. And the photon numbers distributions of Alice's and Bob's sources are
\begin{equation}
a_k^l=\frac{\mu_{l_A}^ke^{-\mu_{l_A}}}{k!},\quad b_k^r=\frac{\mu_{r_B}^ke^{-\mu_{r_B}}}{k!},(l,r=o,x,y,z),
\end{equation}
where $k$ is the photon number in Fock space, and $a_k^l,b_k^r$ are the corresponding probabilities. 

{We denote the two-pulse source as $lr(l,r = o,x,y,z)$ whenever Alice uses her source $l_A$ and Bob uses his source $r_B$. For example, at a certain time window, Alice uses her source $x_A$ and Bob uses his source $x_B$, we shall say that source $xx$ is used at that time window. The time windows with source $zz$ are called signal windows. They are supposed to use effective events caused by single-photon pulse pairs in signal windows to distill the final key through tightened decoy-state analysis.}

\section{The calculation of the final key rate}\label{cal1}
We denote the total number of using instances of source $lr=oo,ox,xo,oy,yo,xy,yx,xx,yy$ as $N_{lr}$, and we have 
\begin{equation}
N_{lr}=p_{l_A}p_{r_B}N.
\end{equation}
According to the data of decoy windows, Alice and Bob get the observed value of the number of effective events of source $lr$, $n_{lr}$. We denote the expected value of $n_{lr}$ as $\mean{n_{lr}}$. We can estimate the lower and upper bounds of $\mean{n_{lr}}$ according to $n_{lr}$ with Chernoff bound which is shown in Appendix \ref{calculation}. And we denote the lower and upper bounds of $\mean{n_{lr}}$ as $\mean{n_{lr}}^L$ and $\mean{n_{lr}}^U$ respectively. Besides, we denote the number of wrong effective events, i.e., wrong bits of source $lr$ as $m_{lr}$ whose corresponding expected value is $\mean{m_{lr}}$. Similarly, we denote the estimated lower and upper bounds of $\mean{m_{lr}}$ as $\mean{m_{lr}}^L$ and $\mean{m_{lr}}^U$ respectively, which can also be estimated by the value of $m_{lr}$ with Chernoff bound. 
\subsection{The prior art results}
To calculate the final key rate of the 4-intensity MDI-QKD, we need to estimate the lower bound of the counting rate and the upper bound of phase-flip error rate of the single-photon pairs in signal windows, $s_{11,Z}^L$ and $e_{11}^{ph,U}$. As shown in Ref.~\cite{zhou2016making}, the expected values of the counting rate and the bit-flip error rate of the single-photon pairs in the decoy windows, $\mean{s_{11,X}}$ and $\mean{e_{11,X}^{bit}}$ satisfy
\begin{equation}
\mean{s_{11,X}}=\mean{s_{11,Z}},\quad \mean{e_{11,X}^{bit}}=\mean{e_{11}^{ph}},
\end{equation}
where $\mean{s_{11,Z}}$ and $\mean{e_{11}^{ph}}$ are the expected values of ${s_{11,Z}}$ and ${e_{11}^{ph}}$. Thus we can first estimate the lower bound of $\mean{s_{11,X}}$ and the upper bound of $\mean{e_{11,X}^{bit}}$ with the data of decoy windows, then we can get the estimated value of $s_{11,Z}^L$ and $e_{11}^{ph,U}$ with Chernoff bound,
\begin{align}
\label{s11z}&s_{11,Z}^L=\frac{O^L(N_{zz}a_1^zb_1^z\mean{s_{11,X}}^L,\xi_{s_{11}})}{N_{zz}a_1^zb_1^z},\\
\label{e11u}&{e_{11}^{ph,U}=\frac{O^U(N_{zz}a_1^zb_1^z s_{11,Z}^L\mean{e_{11,X}^{bit}}^U,\xi_{e_{11}})}{N_{zz}a_1^zb_1^zs_{11,Z}^L},}
\end{align}
where $O^U(Y,\xi)$ and $O^L(Y,\xi)$ are defined in Eqs.~\eqref{OU} and \eqref{OL}. {Here, similar to Ref.~\cite{curty2014finite} we shall use the real values of yield and phase-flip rate of single-photon pulse pairs from signal pulses in calculating the final key rate though their expected values have already taken the major effects of statistical fluctuation~\cite{xu2014protocol,zhou2016making}. Since only those $N_{zz}a_1^zb_1^z s_{11,Z}^L$ untagged bits in the Z basis are valid for the extraction of the final keys, we only need to consider the difference between the expected value and real value of the phase-flip error rate in those untagged bits, thus we get Eq.~\eqref{e11u}.}

 According to the formulas in Ref.~\cite{zhou2016making}, if $\frac{\mu_{y_B}}{\mu_{x_B}}\le\frac{\mu_{y_A}}{\mu_{x_A}}$, we have
\begin{equation}\label{s111}
\mean{s_{11,X}}^L=\frac{\mean{S_+}^L-\mean{S_-}^U-a_1^yb_2^y\mathcal{H}}{a_1^xa_1^y(b_1^xb_2^y-b_2^xb_1^y)},
\end{equation}
where
\begin{align}
&\mean{S_+}=\frac{a_1^yb_2^y}{N_{xx}}\mean{n_{xx}}+\frac{a_1^xb_2^xa_0^y}{N_{oy}}\mean{n_{oy}}+\frac{a_1^xb_2^xb_0^y}{N_{yo}}\mean{n_{yo}},\\
&\mean{S_-}=\frac{a_1^xb_2^x}{N_{yy}}\mean{n_{yy}}+\frac{a_1^xb_2^xa_0^yb_0^y}{N_{oo}}\mean{n_{oo}},\\
\label{mathcalh}&\mathcal{H}=\frac{a_0^x}{N_{ox}}\mean{n_{ox}}+\frac{b_0^x}{N_{xo}}\mean{n_{xo}}-\frac{a_0^xb_0^x}{N_{oo}}\mean{n_{oo}}.
\end{align}
And if $\frac{\mu_{y_B}}{\mu_{x_B}}\ge\frac{\mu_{y_A}}{\mu_{x_A}}$, we have
\begin{equation}\label{s112}
\mean{s_{11,X}}^{L}=\frac{\mean{S_+}^{\prime,L}-\mean{S_-}^{\prime,U}-a_2^yb_1^y\mathcal{H}}{b_1^xb_1^y(a_1^xa_2^y-a_2^xa_1^y)},
\end{equation}
where
\begin{align}
&\mean{S_+}^\prime=\frac{a_2^yb_1^y}{N_{xx}}\mean{n_{xx}}+\frac{a_2^xb_1^xa_0^y}{N_{oy}}\mean{n_{oy}}+\frac{a_2^xb_1^xb_0^y}{N_{yo}}\mean{n_{yo}},\\
&\mean{S_-}^\prime=\frac{a_2^xb_1^x}{N_{yy}}\mean{n_{yy}}+\frac{a_2^xb_1^xa_0^yb_0^y}{N_{oo}}\mean{N_{oo}}.
\end{align}
And the upper bound of $\mean{e_{11,X}^{bit}}$ satisfies 
\begin{equation}\label{e111}
\mean{e_{11,X}^{bit}}^U=\frac{\mean{m_{xx}}^U/N_{xx}-\mathcal{H}/2}{a_1^xb_1^x\mean{s_{11,X}}^L}.
\end{equation}

The derivation details of Eq.~\eqref{e111} are shown in Appendix~\ref{derivation}.

Eqs.~(\ref{s111},\ref{s112},\ref{e111}) are presented by expected values, but we only have observed values from the experiments. We use Chernoff bound to close the gap between the expected values and observed values. To get the tight estimated values of $s_{11,Z}^L$ and $e_{11}^{ph,U}$, we can use the technique of joint constraints~\cite{yu2015statistical}. The details of the analytic results of joint constraints are shown in Sec.~\ref{direct}. Also, we can get the lower and upper bounds of $\mathcal{H}$, $\mathcal{H}^L$ and $\mathcal{H}^U$ with the help of joint constraints. For a certain $\mathcal{H}(\mathcal{H}\in[\mathcal{H}^L,\mathcal{H}^U])$, the key rate~\cite{curty2014finite,zhou2016making} is
\begin{equation}
\begin{split}
R(\mathcal{H})=&p_{z_A}p_{z_B}\{a_1^zb_1^zs_{11,Z}^L[1-h(e_{11}^{ph,U})]-fS_{zz}h(E_{zz})\}\\
&-\frac{1}{N}(\log_2\frac{8}{\varepsilon_{cor}}+2\log_2\frac{2}{\varepsilon^\prime\hat{\varepsilon}}+2\log_2\frac{1}{2\varepsilon_{PA}}),
\end{split}
\end{equation}
where $S_{zz}=n_{zz}/N_{zz}$ is the counting rate of the pulse pairs in signal windows; $E_{zz}$ is the error rate of strings $Z_A$ and $Z_B$; $h(x)=-x\log_2(x)-(1-x)\log_2(1-x)$ is the Shannon entropy; $\varepsilon_{cor}$ is the failure probability of error correction; $\varepsilon_{PA}$ is the failure probability of privacy amplification; and $\varepsilon^\prime$ and $\hat{\varepsilon}$ are the coefficient while using the chain rules of smooth min- and max-entropy.

Finally, by scanning $\mathcal{H}$ in $[\mathcal{H}^L,\mathcal{H}^U]$,  we can get the final key rate
\begin{equation}\label{finalkey}
R=\min_{\mathcal{H}\in[\mathcal{H}^L,\mathcal{H}^U]} R(\mathcal{H}).  
\end{equation}
With the formula in Eq.~\eqref{finalkey}, the total secure coefficient of the 4-intensity MDI-QKD protocol, $\varepsilon_{tol}$ is~\cite{curty2014finite,hayashi2012concise,tomamichel2012tight} 
{
\begin{equation}
\begin{split}
\varepsilon_{tol}=\varepsilon_{cor}+2(\varepsilon^\prime+\hat{\varepsilon}+2\sqrt{\varepsilon_{e}+\varepsilon_{1}})+\varepsilon_{PA},
\end{split}
\end{equation} 
}
where $\varepsilon_{e}$ is the probability that the real value of the phase-flip error rate of the effective events of single photon-pairs in the signal windows is larger than its estimated value $e_{11}^{ph,U}$, and $\varepsilon_{1}$ is the probability that the real value of the counting rate of the single-photon pairs in the signal windows is less than its estimated value $s_{11,Z}^L$.

\subsection{The double-scanning method}
In the original 4-intensity MDI-QKD protocol, an important idea is to consider the constraints jointly. Here, we add new joint constraints with a double-parameter scan: we simultaneously scan the error counts $\mathcal{M}$ and the vacuum related counts $\mathcal{H}$ and get the worst-case jointly for the counting rate of single-photon pulses and phase-flip error rate, where $\mathcal{M}$ is explained below. 

For the effective events of the $xx$ source, they can be divided into two kinds of events, the right effective events and the wrong effective events, which is
\begin{equation}
\mean{n_{xx}}=\mean{\bar{m}_{xx}}+\mean{m_{xx}},
\end{equation}
where $\mean{\bar{m}_{xx}}$ is the expected value of the number of right events of the $xx$ source, and its corresponding observed value is $\bar{m}_{xx}=n_{xx}-m_{xx}$. Denote $\mathcal{M}=\mean{m_{xx}}$.

If $\frac{\mu_{y_B}}{\mu_{x_B}}\le\frac{\mu_{y_A}}{\mu_{x_A}}$, we can rewrite Eq.~\eqref{s111} as
\begin{equation}\label{s113}
\mean{s_{11,X}}^{*L}=\frac{\mean{S_+}^{*L}+\frac{a_1^yb_2^y}{N_{xx}}\mathcal{M}-\mean{S_-}^U-a_1^yb_2^y\mathcal{H}}{a_1^xa_1^y(b_1^xb_2^y-b_2^xb_1^y)},
\end{equation}
where
\begin{align}
&\label{s+*}\mean{S_+}^*=\frac{a_1^yb_2^y}{N_{xx}}\mean{\bar{m}_{xx}}+\frac{a_1^xb_2^xa_0^y}{N_{oy}}\mean{n_{oy}}+\frac{a_1^xb_2^xb_0^y}{N_{yo}}\mean{n_{yo}},\\
\label{s-}&\mean{S_-}=\frac{a_1^xb_2^x}{N_{yy}}\mean{n_{yy}}+\frac{a_1^xb_2^xa_0^yb_0^y}{N_{oo}}\mean{n_{oo}},\\
\label{hhh}&\mathcal{H}=\frac{a_0^x}{N_{ox}}\mean{n_{ox}}+\frac{b_0^x}{N_{xo}}\mean{n_{xo}}-\frac{a_0^xb_0^x}{N_{oo}}\mean{n_{oo}}.
\end{align}
For the case $\frac{\mu_{y_B}}{\mu_{x_B}}\ge\frac{\mu_{y_A}}{\mu_{x_A}}$, we can rewrite Eq.~\eqref{s112} in the similar way.

Then for each group $(\mathcal{H},\mathcal{M})$, we can calculate $s_{11,Z}^{*L}$ and $e_{11}^{ph*,U}$ with Eqs.~(\ref{s11z},\ref{e11u},\ref{s113}) and 
\begin{equation}\label{e1111}
\mean{e_{11,X}^{bit}}^U=\frac{\mathcal{M}/N_{xx}-\mathcal{H}/2}{a_1^xb_1^x\mean{s_{11,X}}^L}.
\end{equation}
Then we have 
\begin{equation}\label{keyimprove}
\begin{split}
&R^*(\mathcal{H},\mathcal{M})=p_{z_A}p_{z_B}\{a_1^zb_1^zs_{11,Z}^{*L}[1-h(e_{11}^{ph*,U})]\\
&-fS_{zz}h(E_{zz})\}-\frac{1}{N}(\log_2\frac{8}{\varepsilon_{cor}}+2\log_2\frac{2}{\varepsilon^\prime\hat{\varepsilon}}+2\log_2\frac{1}{2\varepsilon_{PA}}).
\end{split}
\end{equation}
Finally, by scanning $(\mathcal{H},\mathcal{M})$,  we can get the final key rate
\begin{equation}\label{finalkey2}
\begin{split}
&R^*=\min_{\substack{\mathcal{H}\in[\mathcal{H}^L,\mathcal{H}^U],\\ \mathcal{M}\in [\mathcal{M}^L,\mathcal{M}^U]}} R^*(\mathcal{H},\mathcal{M}),
\end{split}
\end{equation}
whose total secure coefficient $\varepsilon_{tol}^*$ is 
{
\begin{equation}
\begin{split}
\varepsilon_{tol}^*=\varepsilon_{cor}+2(\varepsilon^\prime+\hat{\varepsilon}+2\sqrt{\varepsilon_{e}+\varepsilon_{1}^*})+\varepsilon_{PA},
\end{split}
\end{equation}
}
where $\varepsilon_{1}^*$ is the failure probability that the real value of the counting rate of the single-photon pairs in the signal windows is less than its estimated value $s_{11,Z}^{*L}$.

{In this work we shall use Eq.~\eqref{finalkey2} to calculate the key rate. The key rate can be further improved if we use 
\begin{equation}
R^{*\prime}=R^*+\frac{\underline n_0}{N},
\end{equation}
 where $\underline n_0$ is the lower bound of the number of bits caused by pulse pairs of state $\oprod{0}{0}\otimes \rho_{z_B}$ from source $zz$ and $\underline n_0$ can be verified through observing effective events of source $oz$. Similarly, the key rate of single-scanning method can also be improved if we replace Eq.~\eqref{finalkey} by $R^\prime=R+\underline n_0 /N$.}

\subsection{The analytic results of joint constraints with Chernoff bound}\label{direct}
We shall take Eq.~\eqref{s+*} as an example to show how to get the analytic results of joint constraints with Chernoff bound. To get the lower bound of $\mean{S_+}^*$, we simply replace all the expected values in Eq.~\eqref{s+*} by their estimated lower bounds, which is 
\begin{equation}
\mean{S_+}^{*L}=\frac{a_1^yb_2^y}{N_{xx}}\mean{\bar{m}_{xx}}^L+\frac{a_1^xb_2^xa_0^y}{N_{oy}}\mean{n_{oy}}^L+\frac{a_1^xb_2^xb_0^y}{N_{yo}}\mean{n_{yo}}^L.
\end{equation}
And if the pre-set failure probability by Chernoff bound is $\xi$, the failure probability in estimating $\mean{S_+}^{*}$ is $3\xi$. But if we notice the following joint constraints~\cite{yu2015statistical}
\begin{align*}
&\mean{\bar{m}_{xx}}\ge E^L(\bar{m}_{xx},\xi),\\
&\mean{n_{oy}}\ge E^L(n_{oy},\xi),\\
&\mean{n_{yo}} \ge E^L(n_{yo},\xi),\\
&\mean{\bar{m}_{xx}}+\mean{n_{oy}}\ge E^L(\bar{m}_{xx}+n_{oy},\xi),\\
&\mean{\bar{m}_{xx}}+\mean{n_{yo}}\ge E^L(\bar{m}_{xx}+n_{yo},\xi),\\
&\mean{n_{yo}}+\mean{n_{oy}}\ge E^L(n_{yo}+n_{oy},\xi),\\
&\mean{\bar{m}_{xx}}+\mean{n_{oy}}+\mean{n_{yo}}\ge E^L(\bar{m}_{xx}+n_{oy}+n_{yo},\xi),
\end{align*}
we can apply the technique of linear programming to Eq.~\eqref{s+*} to get better estimated $\mean{S_+}^{*L}$ with those constraints. And at most three of the constrains would be used in the final results, the failure probability in estimating $\mean{S_+}^{*}$ with this method is still $3\xi$. If we run the program of linear programming to solve this problem, much time would be cost especially when we optimize the parameters to get the highest key rate. Fortunately, we have the following analytic results of this special linear programming problem.

We abstract the above linear programming problem into
\begin{align*}
\min_{g_1,g_2,g_3}\quad &F=\gamma_1g_1+\gamma_2g_2+\gamma_3g_3,\\
\mbox{s.t.} \quad &g_1\ge E^L(\widetilde{g}_1,\xi_1),\\
&g_2\ge E^L(\widetilde{g}_2,\xi_1),\\
&g_3\ge E^L(\widetilde{g}_3,\xi_1),\\
&g_1+g_2\ge E^L(\widetilde{g}_1+\widetilde{g}_2,\xi_2),\\
&g_2+g_3\ge E^L(\widetilde{g}_2+\widetilde{g}_3,\xi_2),\\
&g_1+g_3\ge E^L(\widetilde{g}_1+\widetilde{g}_3,\xi_2),\\
&g_1+g_2+g_3\ge E^L(\widetilde{g}_1+\widetilde{g}_2+\widetilde{g}_3,\xi_3),
\end{align*}
where $\gamma_1,\gamma_2,\gamma_3,g_1,g_2,g_3,\widetilde{g}_1,\widetilde{g}_2,\widetilde{g}_3$ all are positive values and $E^L(X,\xi)$ is defined in Eq.~\eqref{EL}. Denoting $\{\gamma_1^*,\gamma_2^*,\gamma_3^*\}$ as the ascending order of $\{\gamma_1,\gamma_2,\gamma_3\}$, and $\{\widetilde{g}_1^*,\widetilde{g}_2^*,\widetilde{g}_3^*\}$ as the corresponding rearrange of $\{\widetilde{g}_1,\widetilde{g}_2,\widetilde{g}_3\}$ according to the ascending order of $\{\gamma_1,\gamma_2,\gamma_3\}$, we have the lower bound of $F$ under those constraints
\begin{equation}\label{fmin}
\begin{split}
&F_L(\gamma_1,\gamma_2,\gamma_3,\widetilde{g}_1,\widetilde{g}_2,\widetilde{g}_3,\xi_1,\xi_2,\xi_3)\\
&=\gamma_1^*E^L(\widetilde{g}_1^*+\widetilde{g}_2^*+\widetilde{g}_3^*,\xi_3)+(\gamma_2^*-\gamma_1^*)E^L(\widetilde{g}_2^*+\widetilde{g}_3^*,\xi_2)\\
&+(\gamma_3^*-\gamma_2^*)E^L(\widetilde{g}_3^*,\xi_1).
\end{split}
\end{equation}
Note that the results of Eq.~\eqref{fmin} may not be the accessible minimum value of the above linear programming problem in some extreme case. But from the perspective of simplifying calculations, we can take  Eq.~\eqref{fmin} as the analytic results and this does not affect the security of the protocol. If we want to get the maximum value under the joint constraints, we can simply replace $E^L(X,\xi)$ by $E^U(X,\xi)$ in Eq.~\eqref{fmin}, where $E^U(X,\xi)$ is defined in Eq.~\eqref{EU}. Specifically, we have the upper bound of $F$ 
\begin{equation}\label{fmax}
\begin{split}
&F_U(\gamma_1,\gamma_2,\gamma_3,\widetilde{g}_1,\widetilde{g}_2,\widetilde{g}_3,\xi_1,\xi_2,\xi_3)\\
&=\gamma_1^*E^U(\widetilde{g}_1^*+\widetilde{g}_2^*+\widetilde{g}_3^*,\xi_3)+(\gamma_2^*-\gamma_1^*)E^U(\widetilde{g}_2^*+\widetilde{g}_3^*,\xi_2)\\
&+(\gamma_3^*-\gamma_2^*)E^U(\widetilde{g}_3^*,\xi_1).
\end{split}
\end{equation}

\section{The optimization method}
To obtain the final key rate with observed values of the experiment, we first calculate the lower bound of $\mean{S_+}^*$ with Eqs.~\eqref{s+*} and \eqref{fmin}, which is
\begin{equation}\label{s++*}
\begin{split}
\mean{S_+}^{*L}=F_L(&\frac{a_1^yb_2^y}{N_{xx}},\frac{a_1^xb_2^xa_0^y}{N_{oy}},\frac{a_1^xb_2^xb_0^y}{N_{yo}},{\bar{m}_{xx}},n_{oy},\\&{n_{yo}},\xi_{S_1^{+*}},\xi_{S_2^{+*}},\xi_{S_3^{+*}}),
\end{split}
\end{equation}
where $\xi_{S_1^{+*}},\xi_{S_2^{+*}},\xi_{S_3^{+*}}$ are the failure probabilities by Chernoff bound, and the following similar symbols are also the failure probabilities. Then we can calculate the upper bound of $\mean{S_-}$ with Eqs.~\eqref{s-} and \eqref{fmax}, which is
\begin{equation}\label{s--*}
\mean{S_-}^U=F_U(\frac{a_1^xb_2^x}{N_{yy}},\frac{a_1^xb_2^xa_0^yb_0^y}{N_{oo}},0,{n_{yy}},{n_{oo}},0,\xi_{S_{1}^-},\xi_{S_2^-},0).
\end{equation}
Similarly, we have the lower and upper bounds of $\mathcal{H}$, which are
\begin{equation}
\begin{split}
\mathcal{H}^L=&F_L(\frac{a_0^x}{N_{ox}},\frac{b_0^x}{N_{xo}},0,n_{ox},n_{xo},0,\xi_{H_1^L},\xi_{H_2^L},0)\\
\\&-\frac{a_0^xb_0^x}{N_{oo}}E^U(n_{oo},\xi_{H_3^L}),\\
\mathcal{H}^U=&F_U(\frac{a_0^x}{N_{ox}},\frac{b_0^x}{N_{xo}},0,n_{ox},n_{xo},0,\xi_{H_1^U},\xi_{H_2^U},0)\\
&-\frac{a_0^xb_0^x}{N_{oo}}E^L(n_{oo},\xi_{H_3^U}).
\end{split}
\end{equation}
It is easy to check that
\begin{equation}
\mathcal{M}^L=E^L(m_{xx},\xi_{m}^L),\quad \mathcal{M}^U=E^U(m_{xx},\xi_{m}^U).
\end{equation}

For each group $(\mathcal{H},\mathcal{M})$, we can calculate the value of $s_{11,Z}^{*L}$ with Eqs.~(\ref{s11z},\ref{s113},\ref{s++*},\ref{s--*}) and the value of $e_{11}^{ph,U}$ with Eqs.~(\ref{e11u},\ref{e1111}). Finally, by scanning $(\mathcal{H},\mathcal{M})$, we obtain the final key rate $R^*$ with Eqs.~\eqref{keyimprove} and \eqref{finalkey2}.

With the calculation method above, the failure probability of the estimation of $s_{11,Z}^{*L}$ is
\begin{equation}
\begin{split}
\varepsilon_{1}^*=&\xi_{S_1^{+*}}+\xi_{S_2^{+*}}+\xi_{S_3^{+*}}+\xi_{S{1}^-}+\xi_{S_2^-}+\xi_{H_1^L}+\xi_{H_2^L}\\
&+\xi_{H_3^L}+\xi_{H_1^U}+\xi_{H_2^U}+\xi_{H_3^U}+\xi_{m}^L+\xi_{m}^U+\xi_{s_{11}},
\end{split}
\end{equation} 
and the failure probability for the estimation of $e_{11}^{ph,U}$ is $\varepsilon_{e}=\xi_{e_{11}}$.

If we set $\varepsilon_{tol}^*$ as a fixed value, then we can regard $R^*$ as the function of those failure probabilities. With the observed values of experiment, we can optimize $R^*$ to get the highest key rates. Besides, in the view of numerical simulation, the observed values could be regarded as the function of source parameters if the channel loss and the properties of detection set-ups are known. That is to say, $R^*$ have the following functional form
\begin{equation}
R^*=R^*(paraA,paraB),
\end{equation}  
where
\begin{align}
paraA=&[p_{x_A},p_{y_A},p_{z_A},\mu_{x_A},\mu_{y_A},\mu_{z_A},p_{x_B},p_{y_B},p_{z_B},\nonumber \\
&\mu_{x_B},\mu_{y_B},\mu_{z_B}],\\
paraB=&[\xi_{S_1^{+*}},\xi_{S_2^{+*}},\xi_{S_3^{+*}},\xi_{S{1}^-},\xi_{S_2^-},\xi_{H_1^L},\xi_{H_2^L},\xi_{s_{11}}\nonumber \\
&\xi_{H_3^L},\xi_{H_1^U},\xi_{H_2^U},\xi_{H_3^U},\xi_{m}^L,\xi_{m}^U,\xi_{e_{11}},\varepsilon_{cor},\varepsilon^\prime,\hat{\varepsilon}].
\end{align}

There are $29$ parameters needed to be optimized if we want to get the highest $R^*$, which is much more than the $6$ parameters in Ref.~\cite{zhou2016making} or $12$ parameters in Ref.~\cite{hu2018measurement}. Thus the optimization method shown in Ref.~\cite{hu2018measurement} does not work well in this optimization problem. In this paper, we would use the random direction method to optimize $R^*$ with $29$ parameters. And the details of the random direction method are shown in Appendix \ref{random}.

\section{Numerical simulation}
\begin{table}[htbp]%\footnotesize
\begin{ruledtabular}
\begin{tabular}{ccccccc}
$p_d$& $e_d$ &$\eta_d$ & $f$ & $\alpha_f$ &$\varepsilon_{tol}$ &$N$\\
\hline
$1.0\times10^{-7}$& $1.5\%$  & $40.0\%$ & $1.1$ & $0.2$ &$1.0\times 10^{-10}$&$1.0\times 10^{10}$\\ 
%%\hline
\end{tabular}
\end{ruledtabular}
\caption{List of experimental parameters used in numerical simulations. Here $p_d$ is the dark counting rate per pulse of Charlie's detectors; $e_d$ is the misalignment-error probability; $\eta_d$ is the detection efficiency of Charlie's detectors; $f$ is the error correction inefficiency; $\alpha_f$ is the fiber loss coefficient ($dB/km$); $\varepsilon_{tol}$ is the total secure coefficient; $N$ is the number of total pulse pairs sent out in the protocol.}\label{exproperty}
\end{table}

We use the linear model to simulate the observed values~\cite{hu2018measurement}. The experimental parameters used in the numerical simulation are listed in Table.~\ref{exproperty}. Without loss of generality, we assume the property of Charlie's detectors are the same. The distance between Alice and Charlie is $L_A$, and that between Bob and Charlie is $L_B$. The total distance between Alice and Bob is $L=L_A+L_B$. In our numerical simulation, we set $L_A=L_B$ for the symmetric case and $L_A-L_B=$constant for the asymmetric case. 

\begin{figure}
\centering
\includegraphics[width=8cm]{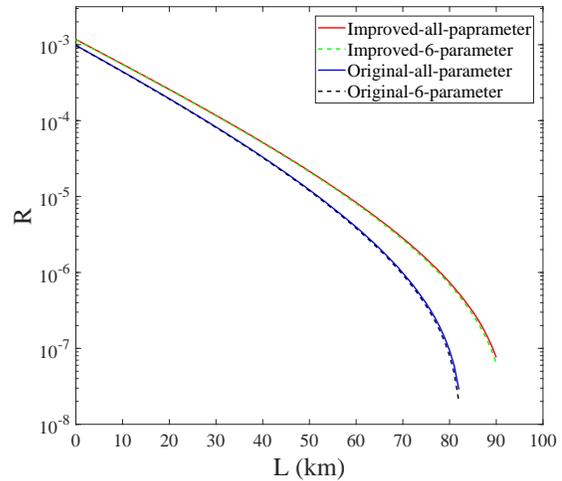}
\caption{Key rates of the 4-intensity protocol through different calculation methods in the symmetric channel. The experimental parameters used here are listed in Table.~\ref{exproperty}. The `Improved-all-parameter' line is the results of double-scanning method of this work optimized with all the parameters including the source parameters and failure probability parameters. The `Improved-6-parameter' line is the results of double-scanning method of this work with only the source parameters optimized. The `Original-all-parameter' line is the results of single-scanning method of Ref.~\cite{zhou2016making} optimized with all the parameters including the source parameters and failure probability parameters. The `Original-6-parameter' line is the results of single-scanning method of Ref.~\cite{zhou2016making} with only the source parameters optimized.}\label{fig3:sy10}
\end{figure}

\begin{figure}
\centering
\includegraphics[width=8cm]{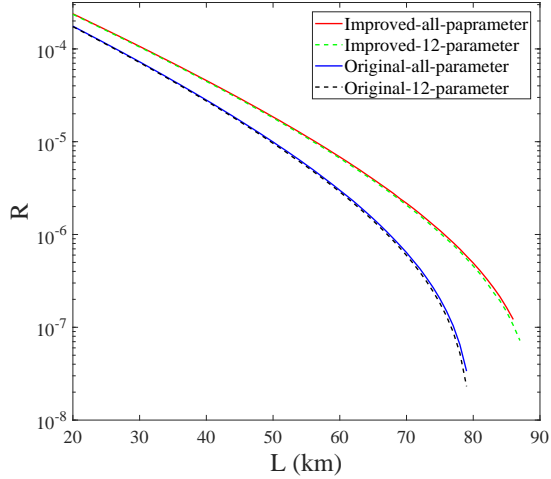}
\caption{Key rates of the 4-intensity protocol through different calculation methods in the asymmetric channel. The distance difference between Alice to Charlie and Bob to Charlie is set as 20 km. The other experimental parameters used here are listed in Table.~\ref{exproperty}. The `Improved-all-parameter' line is the results of double-scanning method of this work optimized with all the parameters including the source parameters and failure probability parameters. The `Improved-12-parameter' line is the results of double-scanning method of this work with only the source parameters optimized. The `Original-all-parameter' line is the results of single-scanning method of Ref.~\cite{zhou2016making} optimized with all the parameters including the source parameters and failure probability parameters. The `Original-12-parameter' line is the results of single-scanning method of Ref.~\cite{zhou2016making} with only the source parameters optimized.}\label{fig3:asy10}
\end{figure}

Figure \ref{fig3:sy10} and Figure \ref{fig3:asy10} are the numerical results of this work and the original 4-intensity MDI-QKD protocol with the symmetric channel and asymmetric channel, respectively. Only in the symmetric case, we set symmetric source parameters for Alice and Bob, that is to say, $p_{x_A}=p_{x_B}$, $\mu_{x_A}=\mu_{x_B}$ and so on. {If the channel is asymmetric, we do not take any of this setting. In the asymmetric case, we assume $p_{\gamma_A}\neq p_{\gamma_B}$ and $\mu_{\gamma_A}\neq \mu_{\gamma_B}$ for $\gamma=o,x,y,z$.} In the asymmetric case, the distance difference between Alice to Charlie and Bob to Charlie is set as 20 km. The `Improved-all-parameter' line is the results of double-scanning method of this work optimized with all the parameters including the source parameters and failure probability parameters. The `Improved-6-parameter' or `Improved-12-parameter' line is the results of double-scanning method of this work with only the source parameters optimized. The `Original-all-parameter' line is the results of single-scanning method of Ref.~\cite{zhou2016making} optimized with all the parameters including the source parameters and failure probability parameters. The `Original-6-parameter' or `Original-12-parameter' line is the results of single-scanning method with only the source parameters optimized. The simulation results show that the method in this work can improve the key rate of 4-intensity MDI-QKD protocol, especially when the channel loss is large. The simulation results show that the optimized results with all parameters is almost the same as the optimized results with only sources parameters.

\begin{table*}
	\centering
	
	\begin{tabular}{ccccccc}
		\hline
		\multirow{2}*{\diagbox[width=6em]{method}{$L$}}&\multicolumn{2}{c}{25 km}&\multicolumn{2}{c}{50 km}&\multicolumn{2}{c}{75 km}\\
		 &SPO& APO&SPO&APO&SPO&APO\\
		\hline
		Ref.~\cite{zhou2016making} & $1.26\times 10^{-4}$& $1.27\times 10^{-4}$ &$1.19\times 10^{-5}$ & $1.22\times 10^{-5}$ &$3.61\times 10^{-7}$& $3.92\times 10^{-7}$\\
		 This work & $1.72\times 10^{-4}$& $1.74\times 10^{-4}$ &$2.11\times 10^{-5}$ & $2.15\times 10^{-5}$ &$1.45\times 10^{-6}$& $1.52\times 10^{-6}$\\
		\hline
	\end{tabular}
	\caption{The key rates of this work and the original 4-intensity MDI-QKD protocol in the symmetric channel. The experimental parameters used here are listed in Table.~\ref{exproperty}. SPO: source-parameter-optimization; APO: all-parameter-optimization.}\label{tab:cha3improvecom}
\end{table*}

Table \ref{tab:cha3improvecom} is the comparison of the key rates of this work and the original 4-intensity MDI-QKD protocol in the symmetric channel. The experimental parameters used here are listed in Table.~\ref{exproperty}. {Here we take the distances of 25 km, 50 km and 75 km as examples to show the improvement of our new method. The results show that as the distance increases, the influence of statistical fluctuations becomes more and more significant, and the key rate advantage of our method is also increasing. For the distance of 25 km, our method improves the key rate by 35\%. And for the distance of 75 km, our method improves the key rate by 280\%.} Compared with the optimized results with only the source parameters, the all-parameter optimize method could improve the key rate by about $10\%$.

\begin{figure}
\centering
\includegraphics[width=8cm]{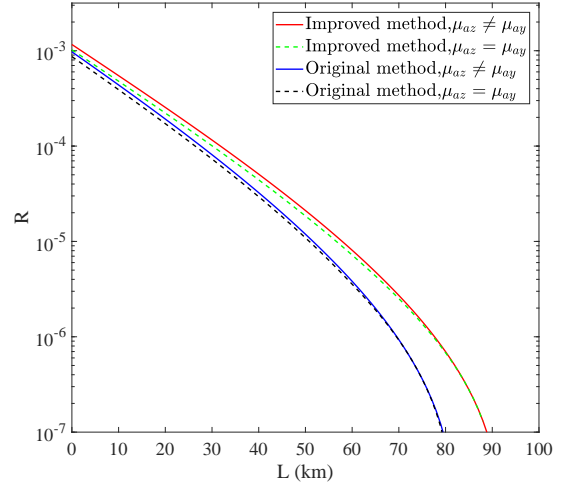}
\caption{ Key rates of the 4-intensity protocol and 3-intensity protocol through different calculation methods in the symmetric channel. The experimental parameters used here are listed in Table. I. The `Improved method' lines are the results of the double-scanning method, and the “Original method” lines are the results of the single scanning method [46]. The “$\mu_{az}\neq \mu_{ay}$” lines are the results of the 4-intensity protocol and the “$\mu_{az}= \mu_{ay}$” lines are the results of the 3-intensity protocol.}\label{com34}
\end{figure}

{\textbf{The unbalanced 3-intensity protocol and its performance.} In the special case of $\mu_{z_A}=\mu_{y_A}$ and $\mu_{z_B}=\mu_{y_B}$, our 4-intensity protocol becomes the unbalanced 3-intensity protocol. Numerical simulation shows that the unbalanced 3-intensity  protocol with double-scanning method can also present quite good key rate which is even better than the 4-intensity protocol with single-scanning method. See in Figure~\ref{com34} for details.
}

\section{Conclusion}
Based on the 4-intensity MDI-QKD protocol, we propose a double-scanning method to further improve the key rate. Numerical results show that the method in this work can improve the key rate by $35\%-280\%$. The method in this work can directly apply to the existing experiments.

\appendix
\section{Chernoff bound}\label{calculation}
The Chernoff bound can help us estimate the expected value from their observed values~\cite{jiang2017measurement,chernoff1952measure}. Let $X_1,X_2,\dots,X_n$ be $n$ random samples, detected with the value 1 or 0, and let $X$ denote their sum satisfying $X=\sum_{i=1}^nX_i$. $E$ is the expected value of $X$. We have
\begin{align}
\label{EL}E^L(X,\xi)=&\frac{X}{1+\delta_1(X,\xi)},\\
\label{EU}E^U(X,\xi)=&\frac{X}{1-\delta_2(X,\xi)},
\end{align}
where we can obtain the values of $\delta_1(X)$ and $\delta_2(X)$ by solving the following equations
\begin{align}
\label{delta1}\left(\frac{e^{\delta_1}}{(1+\delta_1)^{1+\delta_1}}\right)^{\frac{X}{1+\delta_1}}&=\xi,\\
\label{delta2}\left(\frac{e^{-\delta_2}}{(1-\delta_2)^{1-\delta_2}}\right)^{\frac{X}{1-\delta_2}}&=\xi,
\end{align}
where $\xi$ is the failure probability.

Moreover, we can use the Chernoff bound to help us estimate their real values from their expected values. Similar to Eqs.~\eqref{EL}- \eqref{delta2}, the observed value, $O$, and its expected value, $Y$, satisfy
\begin{align}
\label{OU}&O^U(Y,\xi)=[1+\delta_1^\prime(Y,\xi)]Y,\\
\label{OL}&O^L(Y,\xi)=[1-\delta_2^\prime(Y,\xi)]Y,
\end{align}   
where we can obtain the values of $\delta_1^\prime(Y,\xi)$ and $\delta_2^\prime(Y,\xi)$ by solving the following equations
\begin{align}
\left(\frac{e^{\delta_1^\prime}}{(1+\delta_1^\prime)^{1+\delta_1^\prime}}\right)^{Y}&=\xi,\\
\label{endd}\left(\frac{e^{-\delta_2^\prime}}{(1-\delta_2^\prime)^{1-\delta_2^\prime}}\right)^{Y}&=\xi.
\end{align}

\section{The random direction method}\label{random}
\textbf{Initialization} Find a original point $Para=[paraA_o,paraB_o]$ where $R^*(Para)> 0$. Set initial step $d_{step}$ and minimum step $d_{min}$. Set the maximum number of cycles $C_{max}$.

\textbf{(i). } If $d_{step}<d_{min}$, stop the optimization programme and output the value of $R_{opt}=R^*(Para)$ as the optimal key rate, where $Para$ is the corresponding optimal parameters; If $d_{step}>d_{min}$, set the cycle count $C=1$. Then go to step (ii).

\textbf{(ii). } If $C>C_{max}$, let $d_{step}:=d_{step}/5$, then go to step (i); if $C\le C_{max}$, go to step (iii).

\textbf{(iii). } Use a Gaussian random number generator to generate 29 random numbers, then normalize these random numbers and put them into the array $D_{dir}$. Then calculate $R_{temp} = R_f^* (Para + d_{step} \times D_{dir})$. If $R_{temp}> R_{opt}$, then let $Para: = Para + d_{step} \times D_{dir}$,  $R_{opt} = R_{temp}$, $C = 1$; if $R_{temp} \le R_{opt}$, then let $C: = C + 1$. Finally go to step (ii).

\section{The derivation of Eq.~\eqref{e111}}\label{derivation}
We have the following definitions of set $\mathcal{C}$ and $c_{jk}$ to clearly show our derivation process. 

Definition. Set $\mathcal{C}$ and $c_{jk}$: set $\mathcal{C}$ contains all those effective wrong bits caused by non-vacuum $\ket{jk}$($j \ge 1,k \ge 1$)-photon pulse pairs from source $lr(l,r = o,x,y)$ and all those effective bits caused by vacuum $\ket{jk}$($j=0$ or $k=0$)-photon pulse pairs of sources $lr(l, r = o, x, y)$. Set $c_{jk}$ contains all bits caused by $\ket{jk}$-photon pulse pairs in set $\mathcal{C}$. Obviously, for those bits in set $\mathcal{C}$ caused by vacuum pulse pairs (the pulse pair including at least one vacuum pulse), everyone has an independent probability $\frac{1}{2}$ to be a wrong bit. We shall use this important fact.

For any bit $i$, if $i\in c_{jk}$, the probability that it is a bit from source $lr$ is 
\begin{equation}\label{qijk}
Q_{jk}^{lr}=p_{l_A}p_{r_B}a_j^lb_k^r d_{jk},
\end{equation}
where
\begin{equation}
d_{jk}=\frac{1}{\sum_{l,r=o,x,y}p_{l_A}p_{r_B}a_j^lb_k^r}.
\end{equation}

Also, for any bit $i$, if $i\in c_{jk}$, the probability that it is a wrong bit from source $lr$ is 
\begin{equation}\label{pijk}
P_{jk}^{lr}=\left\{
\begin{split}
Q_{jk}^{lr}, \mbox{ if } j\ge 1\mbox{ and } k\ge 1,\\
\frac{1}{2}Q_{jk}^{lr}, \mbox{ if } j=0\mbox{ or } k=0.
\end{split}
\right.
\end{equation}

We define
\begin{equation}\label{mjklr}
\mean{m_{jk}^{lr}}=\sum_{i\in c_{jk}}P_{jk}^{lr},
\end{equation}
which is the expected value of the number of wrong bits caused by $\ket{jk}$-photon pulse-pair from source $lr$. Mathematically, we also define
\begin{equation}\label{meanmlr}
\mean{m_{lr}}=\sum_{j\ge 0,k\ge 0}\mean{m_{jk}^{lr}},
\end{equation}
where $\mean{m_{lr}}$ is called the expected value of the number of wrong bits caused by source $lr$.

Specifically, the expected value of the number of wrong bits caused by source $xx$ is $\mean{m_{xx}}$ and
\begin{equation}\label{eqmxx1}
\begin{split}
\mean{m_{xx}}=&\mean{m_{11}^{xx}}+\sum_{j\ge 1,k\ge 1,jk\neq 1}\mean{m_{jk}^{xx}}\\
&+\sum_{j\ge 0}\mean{m_{j0}^{xx}}+\sum_{k\ge 0}\mean{m_{0k}^{xx}}-\mean{m_{00}^{xx}}.
\end{split}
\end{equation}
Therefore the expected value of the number of wrong bits caused by single-photon pulse pairs from source $xx$ respects the following inequality
\begin{equation}\label{eqmxx11}
\begin{split}
\mean{m_{11}^{xx}}\le \mean{m_{xx}}-\sum_{j\ge 0}\mean{m_{j0}^{xx}}-\sum_{k\ge 0}\mean{m_{0k}^{xx}}+\mean{m_{00}^{xx}}.
\end{split}
\end{equation}

On the other hand, using Eqs.~\eqref{qijk} and \eqref{pijk}, we have other formulas for $\mean{m_{xx}}$, the expected number of wrong bits from source $xx$.

According to Eqs.~\eqref{qijk} and \eqref{pijk}, for any bit $i \in \mathcal{C}$, it has a probability $p_i$ to be a wrong bit caused by source $xx$, and 
\begin{equation}
p_i=\left\{
\begin{split}
p_{x_A}p_{x_B}a_j^xb_k^x d_{jk}, \mbox{ if } i\in c_{jk} \mbox{ and }j\ge 1\mbox{ and } k\ge 1,\\
\frac{1}{2}p_{x_A}p_{x_B}a_j^xb_k^x d_{jk},  \mbox{ if } i\in c_{jk} \mbox{ and }j=0\mbox{ or } k=0.
\end{split}
\right.
\end{equation}
We rewrite Eq.~\eqref{eqmxx1} as
\begin{equation}
\mean{m_{xx}}=\sum_{i\in\mathcal{C}}p_i,
\end{equation}
where all values of $\{p_i\}$ are independent. 

Since all $p_i$ here are independent, we can bound this expected value based on the observed value by Chernoff bound, say 
\begin{equation}
\mean{m_{xx}}\le \mean{m_{xx}}^U=\frac{m_{xx}}{1-\delta_{xx}},
\end{equation}
where $\delta_{xx}$ can be calculated by Eq.~\eqref{delta1}

%Besides, We define
%\begin{equation}\label{mjklr}
%\mean{n_{jk}^{lr}}=\sum_{i\in c_{jk}}Q_{jk}^{lr},j=0 \mbox{ or } k=0,
%\end{equation}
%which is the expected value of the number of bits caused by $\ket{jk}$($j=0$ or $k=0$)-photon pulse-pair from source $lr$. Mathematically, we also define
%\begin{equation}\label{meanmlr}
%\mean{n_{lr}}=\sum_{j\ge 0,k\ge 0}\mean{n_{jk}^{lr}}, l=o \mbox{ or } r=o,
%\end{equation}
%where $\mean{n_{lr}}$ is called the expected value of the number of wrong bits caused by source $lr$($l=o$ or $r=o$).

Similarly, we can also rewrite the term $\sum_{j\ge 0}\mean{m_{j0}^{xx}}+\sum_{k\ge 0}\mean{m_{0k}^{xx}}-\mean{m_{00}^{xx}}$ in Eq.~\eqref{eqmxx11} so as to relate the terms with the expected values of the number of bits caused by sources $xo,ox,oo$, and therefore finally relate them to the scanning parameter $\mathcal{H}$. For example, to the term $\sum_{k\ge 0}\mean{m_{0k}^{xx}}$, we first consider sets ${c_{0k}}$ for all $k$. To any bit $i\in c_{0k}$, it has a probability $P_k=\frac{1}{2}p_{x_A}p_{x_B}a_0^xb_k^x d_{0k}$ to be a wrong bit caused by $xx$ source, that is 
\begin{equation}\label{kge}
\sum_{k\ge 0}\mean{m_{0k}^{xx}}=\sum_{k\ge 0}\sum_{i\in c_{0k}}P_k=\sum_{k\ge 0}N_{c_{0k}}P_k,
\end{equation}
where $N_{c_{0k}}$ is the number of elements in $c_{0k}$. Also, to any bit $i\in c_{0k}$, it has a probability $Q_k=p_{o_A}p_{x_B}b_k^x d_{0k}$ to be a bit caused by $ox$ source, which means
\begin{equation}\label{nox}
\mean{n_{ox}}=\sum_{k\ge 0}\sum_{i\in c_{0k}}Q_k=\sum_{k\ge 0}N_{c_{0k}}Q_k,
\end{equation} 
where $\mean{n_{ox}}$ is the expected value of the number of bits caused by sources $ox$. Since $\frac{P_k}{Q_k}=\frac{p_{x_A}a_{0}^x}{2p_{o_A}}$ holds for all $k$, comparing Eqs.~\eqref{kge} and \eqref{nox}, we have
\begin{equation}
\sum_{k\ge 0}\mean{m_{0k}^{xx}}=\frac{p_{x_A}a_{0}^x}{2p_{o_A}}\mean{n_{ox}}.
\end{equation}

In a similar way, we also have 
\begin{equation}
\sum_{j\ge 0}\mean{m_{j0}^{xx}}=\frac{p_{x_B}b_{0}^x}{2p_{o_B}}\mean{n_{xo}}, \mean{m_{00}^{xx}}=\frac{p_{x_A}p_{x_B}a_0^xb_0^x}{2p_{o_B}}\mean{n_{oo}}, 
\end{equation}
where $\mean{n_{xo}}$ and $\mean{n_{oo}}$ are the expected values of the number of bits caused by sources $xo,oo$ respectively.

With these, Eq.~\eqref{eqmxx11} is changed into
\begin{equation}
\begin{split}
\mean{m_{11}^{xx}}\le &\mean{m_{xx}}-\frac{p_{x_A}a_0^x}{2p_{o_A}}\mean{n_{ox}}-\frac{p_{x_B}b_0^x}{2p_{o_B}}\mean{n_{xo}}\\
&+\frac{p_{x_A}p_{x_B}a_0^xb_0^x}{2p_{o_B}}\mean{n_{oo}}.
\end{split}
\end{equation}

We define
\begin{equation}
\mean{e_{11,X}^{bit}}=\frac{\mean{m_{11}^{xx}}}{Np_{x_A}p_{x_B}a_1^xb_1^x\mean{s_{11,X}}}.
\end{equation}
According to the definition of $\mean{m_{11}^{xx}}$ by Eq.~\eqref{mjklr}, this is the bit-flip error rate of all those single-photon pairs in the $X$ basis and this can be used to estimate the phase-flip rate of single-photon pulses in the $Z$ basis. Recall the definition of $\mathcal{H}$ in Eq.~\eqref{mathcalh}, we have
\begin{equation}
\mean{e_{11,X}^{bit}}^U=\frac{\mean{m_{xx}}^U/N_{xx}-\mathcal{H}/2}{a_1^xb_1^x\mean{s_{11,X}}^L}.
\end{equation}

This ends the proof of Eq.~\eqref{e111}.

\bibliography{refs}

\end{document}